\documentclass[fleqn,12pt,twoside]{article}
\usepackage{espcrc1}

\usepackage{graphicx}
\usepackage[figuresright]{rotating}

\newcommand{\AmS}{{\protect\the\textfont2J.P. Ralston and B. Pire,
  A\kern-.1667em\lower.5ex\hbox{M}\kern-.125emS}}
\newcommand{\be}{\begin{eqnarray}}
\newcommand{\ee}{\end{eqnarray}}
\newcommand{\bea}{\begin{eqnarray}}
\newcommand{\eea}{\end{eqnarray}}
\title{Position space interpretation for generalized
parton distributions}

\author{Matthias Burkardt \address[MCSD]{
Department of Physics, New Mexico State University,
Las Cruces, NM 88011, U.S.A.}
        \thanks{This work was supported by the DOE
(DE-FG03-95ER40965)} }
       
\begin{document}

% typeset front matter
\maketitle

\begin{abstract}
For an unpolarized target,
the generalized parton distribution $H_q(x,0,t)$ 
is related to the distribution of partons in impact 
parameter space. 
The transverse distortion of this distribution
for a transversely polarized target is described
by $E_q(x,0,t)$. 
\end{abstract}

\section{INTRODUCTION}
Generalized parton distributions (GPDs)
\cite{gpd} have
attracted significant interest since it has been
recognized that they can not only be probed in
deeply virtual Compton scattering experiments but
can also be related to the orbital angular momentum
carried by quarks in the nucleon \cite{xdj}.
However, remarkably little is still known about
the physical interpretation of GPDs, and one may
ask the question: {\it
suppose, about 10-15 years from now, after a 
combined effort 
from experiment, simulation and theory, we know 
how these functions (i.e. GPDs) look like
for the nucleon. What is it, in simple physical 
terms, that we will have learned about the structure
of the nucleon?} Of course, we will have learned 
something about the orbital angular momentum
carried by the quarks \cite{xdj}, but is that all 
there is?
In these notes, I will discuss another interesting
piece of information that can be extracted from
GPDs, namely {\it how partons are distributed in the
transverse plane}. 

In nonrelativistic quantum mechanics, the physics
of form factors is illucidated by transforming to
the center of mass frame and by interpreting the
Fourier transform of form factors as charge 
distributions in that frame.

GPDs \cite{gpd} are the form factors of the same 
operators [light cone correlators $\hat{O}_q(x,{\bf 0_\perp})$]
whose forward matrix elements also yield the usual
(forward) parton distribution functions (PDFs).
For example, the unpolarized PDF $q(x)$ can be
expressed in the form\footnote{%Throughout this work 
We will suppress the scale (i.e. $Q^2$)
dependence of these matrix elements for notational
convenience. In the end, the $\perp$ `resolution' will
be limited by $1/Q$.}
\be 
q(x) = \langle p,\lambda|
\hat{O}_q(x,{\bf 0_\perp})
|p,\lambda \rangle ,
\label{eq:pdf}
\ee
while the GPDs $H_q(x,\xi,t)$ and $E_q(x,\xi,t)$ are defined as
\be
\langle p^\prime,\lambda^\prime |
\hat{O}_q(x,{\bf 0_\perp})
|p,\lambda \rangle 
= \frac{1}{2\bar{p}^+}
\bar{u}(p^\prime,\lambda^\prime)\left(\gamma^+  
H_q(x,\xi,t)
+ i\frac{\sigma^{+\nu}\Delta_\nu}{2M} E_q(x,\xi,t)
\right)u(p,\lambda) \label{eq:gpd},
\ee
where $\Delta=p^\prime -p$, $2\bar{p}= p+p^\prime$,
$t=\Delta^2$, $2\bar{p}^+\xi = \Delta^+$, and
\be
\hat{O}_q(x,{\bf b_\perp}) = 
\int \frac{dx^-}{4\pi}\,\bar{q}
\left(-\frac{x^-}{2},{\bf b_\perp} \right) \gamma^+ 
q\left(\frac{x^-}{2},{\bf b_\perp}\right) 
e^{ixp^+x^-}.
\label{eq:0perp}
\ee 
In the case of form factors, 
non-forward matrix elements provide information
about how the charge (i.e. the
forward matrix element) is distributed in
position space.
By analogy with form factors, one would therefore
expect that the additional 
information (compared to PDFs)
contained in GPDs helps to understand how
the usual PDFs are distributed in position space
\cite{ralston}.
Of course, since the operator 
$\hat{O}_q(x,{\bf 0_\perp})$
already `filters out' quarks with a definite
momentum fraction $x$, Heisenberg's uncertainty
principle does not allow a simultaneous determination
of the partons' longitudinal position, but
determining the distributions of partons in
impact parameter space is conceiveable.
Making these intuitive expectation more precise
(e.g. what is the `reference point', `are there
relativistic corrections', `how
does polarization enter', `is there a strict
probability interpretation')
will be the main purpose of these notes. 

\section{IMPACT PARAMETER DEPENDENT PARTON
DISTRIBUTIONS}

In nonrelativistic quantum mechanics, the Fourier
transform of the form factor yields the charge
distribution in the center of mass frame.
In general, the concept of a center of mass 
has no analog in relativistic theories, and 
therefore the position space interpretation of form
factors is frame dependent.

The infinite momentum frame (IMF) plays a
distinguished role in the physical interpretation
of regular PDFs as momentum distributions in the IMF.
It is therefore natural to attempt to interpret
GPDs in the IMF. This task is facilitated by the
fact  that there a is Galilean subgroup
of transverse boosts in the IMF, whose generators
are defined as
\be
B_x \equiv M^{+x} =
{(K_x+J_y)}/{\sqrt{2}} 
\quad\quad\quad\quad 
B_y \equiv
M^{+y} = {(K_y-J_x)}/{\sqrt{2}} ,
\ee
where $M_{ij}=\varepsilon_{ijk}J_k$,
$M_{i0}=K_i$, and $M^{\mu \nu}$ is the familiar 
generator of Lorentz transformations.  
The commutation relations between ${\bf B_\perp}$
and other Poincar\'e generators
\be
\left[J_3,B_k\right]&= i\varepsilon_{kl}B_l
\quad\quad\quad\quad 
\left[P_k,B_l\right] &= -i\delta_{kl}P^+
\nonumber\\
\left[P^-,B_k\right] &= -iP_k 
\quad\quad\quad\quad
\left[P^+,B_k\right]&= 0
\ee
where $k,l\in\{x,y\}$, $\varepsilon_{xy}=-
\varepsilon_{yx}=1$, and $\varepsilon_{xx}=
\varepsilon_{yy}=0$, are formally identical
to the commutation relations among 
boosts/translations for a nonrelativistic system in 
the plane provided we make the identification
\cite{soper}
\be
\begin{array}{ll}
{\bf P}_\perp \longrightarrow \mbox{momentum in the
plane}
& P^+ \longrightarrow \mbox{mass}
\\
L_z\longrightarrow \mbox{rotations around $z$-axis}
%\nonumber\\
&P^-\longrightarrow \mbox{Hamiltonian}\\
{\bf B_\perp} \longrightarrow
\mbox{generator of boosts in the $\perp$ plane}.&
\end{array} 
\ee	
Because of this isomorphism it is possible to
transfer a number of results and concepts from
nonrelativistic quantum mechanics to the infinite
momentum frame. For example, for an eigenstate of 
$P^+$, one can define a (transverse) 
{\it center of momentum} (CM)
\be
{\bf R_\perp} \equiv - \frac{{\bf B_\perp}}{p^+}
= \int dx^- \int d^2{\bf x_\perp}
T^{++} {\bf x_\perp},
\label{eq:Rperp}
\ee
where $T^{\mu \nu}$ is the energy momentum tensor.
Like its nonrelativistic counterpart, it satisfies
$\left[J_3,R_k\right]= i\varepsilon_{kl}R_l$
and
$\left[P_k,R_l\right]= -i\delta_{kl}$. These
simple commutation relations enable us to form 
simultaneous eigenstates of ${\bf R_\perp}$ 
(with eigenvalue ${\bf 0_\perp}$), $P^+$ and $J_3$
\be
\left|p^+,{\bf R_\perp}={\bf 0_\perp},\lambda
\right\rangle
\equiv {\cal N}\int d^2{\bf p_\perp}
\left|p^+,{\bf p_\perp},\lambda
\right\rangle,
\label{eq:perpcm}
\ee
where ${\cal N}$ is some normalization constant,
and $\lambda$ corresponds to the helicity
when viewed from a frame with infinite momentum. 
For details on how these IMF helicity states
are defined, as well as for their relation to
usual rest frame states, see Ref. \cite{soper1}.

In the following we will use the eigenstates of the 
$\perp$ center of momentum operator\footnote{Note 
that the Galilei invariance in the IMF is 
crucial for being able to construct a useful
CM concept.} (\ref{eq:perpcm}) to 
define the concept of a parton distributions in
impact parameter space
\footnote{In Ref. \cite{mb1}, wave packets were used
in order to avoid states that are normalized to 
$\delta$ functions. The final results are unchanged.
This was also verified in Ref. \cite{diehl2}.}
\be
q(x,{\bf b_\perp})\equiv
\left\langle p^+,{\bf R_\perp}={\bf 0_\perp},\lambda
\right|\hat{O}_q(x,{\bf b_\perp})
\left|p^+,{\bf R_\perp}={\bf 0_\perp},\lambda 
\right\rangle.
\label{eq:bPDF}
\ee
It is straightforward to verify that the impact 
parameter dependent PDFs defined above 
(\ref{eq:bPDF}) are the Fourier transform of
$H_q$ \cite{mb1,mb2,diehl2} (without
relativistic corrections!)
\footnote{
A similar interpretation exists for 
$\tilde{H}_q(x,0,t)$ in terms of impact parameter
dependent polarized quark distributions
$
\Delta q(x,{\bf b_\perp}) = 
\int \frac{d^2{\bf \Delta_\perp}}{(2\pi)^2}
\tilde{H}_q(x,0,-{\bf \Delta}^2_\perp)
e^{i{\bf \Delta_\perp}\cdot {\bf b_\perp}}.
$}
\be 
q(x,{\bf b_\perp})\!\!\!\!
&=& \!\!\!\!\frac{\left|{\cal N}\right|^2}{(2\pi)^2}
\! \int \!\!d^2{\bf p}_\perp\!  \int \!\!
d^2{\bf p}_\perp^\prime
\left\langle p^+,{\bf 0_\perp},\lambda\left|
\hat{O}_q(x,{\bf b_\perp})
\right|p^+,{\bf 0_\perp},\lambda\right\rangle
\label{eq:result1}\\
\!\!\!\!&=&\!\!\!\!
\frac{\left|{\cal N}\right|^2}{(2\pi)^2}\!
 \int\!\! d^2{\bf p}_\perp \! \int \!\!
d^2{\bf p}_\perp^\prime
H_q(x,-\left({\bf p}_\perp^\prime\!-\!
{\bf p}_\perp
\right)^2) 
e^{i{\bf b_\perp}
\cdot ({\bf p}_\perp^\prime\!-\!{\bf p}_\perp)}
%\nonumber\\
=\!\!\!
 \int \!\!\frac{d^2{\bf \Delta}_\perp}{(2\pi)^2}  
H_q(x,\!-{\bf \Delta}_\perp^2) e^{i{\bf b_\perp} \cdot
{\bf \Delta}_\perp}\!\!\!
\nonumber
\ee
and its normalization is $\int d^2{\bf b}_\perp
q(x,{\bf b_\perp})=q(x)$.
Furthermore, $q(x,{\bf b_\perp})$ has a probabilistic
interpretation. Denoting $\tilde{b}_s(k^+,
{\bf b_\perp})$ [$\tilde{d}_s(k^+,
{\bf b_\perp})$] the canonical destruction operator
for a quark [antiquark]
with longitudinal momentum $k^+$ and
$\perp$ position ${\bf b_\perp}$, one finds 
\cite{mb2}
\bea
q(x,{\bf b_\perp})=\left\{
\begin{array}{ll}\sum_s
\left|\tilde{b}_s(xp^+,
{\bf b_\perp})\left|p^+,{\bf 0_\perp},\lambda
\right\rangle \right|^2\geq 0&
\quad \quad \quad \mbox{for}\quad x>0\\
\sum_s
\left|\tilde{d}_s^\dagger(xp^+,
{\bf b_\perp})\left|p^+,{\bf 0_\perp},\lambda
\right\rangle \right|^2\leq 0&
\quad \quad \quad \mbox{for}\quad x<0\end{array}
\right.
\eea
For large $x$, one expects $q(x,{\bf b_\perp})$ 
to be not only small in magnitude (since $q(x)$
is small for large $x$) but also very narrow
(localized valence core!). In particular, the
$\perp$ width should vanish as $x\rightarrow 1$,
since $q(x,{\bf b_\perp})$ is defined with the
$\perp$ CM as a reference point.
A parton representation for ${\bf R_\perp}$ 
(\ref{eq:Rperp}) is given by
$%\be
{\bf R_\perp}=\sum_{i\in q,g} x_i {\bf r_{i,\perp}},
$%\ee
where $x_i$ (${\bf r_{i,\perp}}$) is the momentum
fraction ($\perp$ position) of the $i^{th}$ parton,
and for $x=1$ the position of active quark 
coincides with the $\perp$ CM.

\begin{figure}
\begin{picture}(100,235)(140,710)
\includegraphics{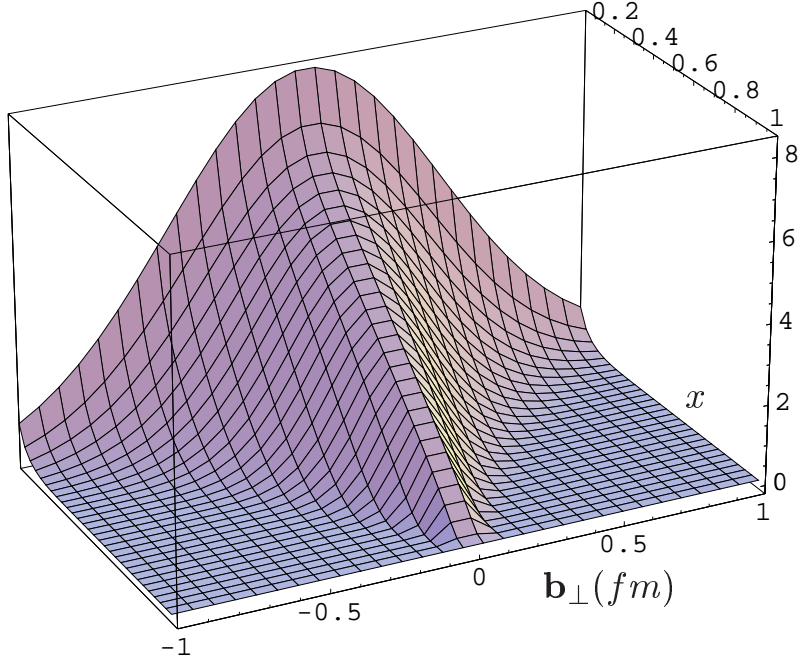}
\end{picture}
\caption{Impact parameter dependent parton
distribution $q(x,{\bf b_\perp})$ for the
model (\ref{eq:model}).}
\label{fig:model}
\end{figure}

In order to gain some intuition for the kind
of results that one might expect for impact parameter
dependent PDFs, we consider a simple model
\be
H_q(x,0,-{\bf \Delta}_\perp^2) = q(x)
e^{-a{\bf \Delta}_\perp^2(1-x)\ln \frac{1}{x}}.
\label{eq:log2}
\ee
The precise functional form in this ansatz should
not be taken too seriously, and the model should
only be considered a simple parameterization
which is consistent with both Regge behavior
at small $x$ and 
Drell-Yan-West duality at large $x$. A 
straightforward Fourier transform yields
(Fig. \ref{fig:model})
\be
q(x,{\bf b_\perp}) = q(x)\frac{1}{4\pi(1-x)\ln
\frac{1}{x}}
e^{-\frac{{\bf b_\perp}^2}{4a(1-x)\ln\frac{1}{x}}}.
\label{eq:model}
\ee

\section{POSITION SPACE INTERPRETATION FOR
$E(x,0,-{\bf \Delta_\perp^2})$}
While both $H(x,0,t)$ and $\tilde{H}(x,0,t)$ are
diagonal in helicity, $E(x,0,t)$ contributes only
to helicity flip matrix elements. In fact 
for $p^+=p^{+\prime}$ (i.e. $\xi=0$) \cite{flip}
\cite{diehl}
\bea
\int \frac{dx^-}{4\pi} e^{ip^+x^- x}
\left\langle P\!\!+\!\!\Delta,\! \uparrow\!
\left| \bar{q}\!\left(\!0\!\right)\gamma^+
q\!\left(\!{x^-}\!\right)
\right| P,\! \uparrow \right\rangle
&=&H(x,\!0,\!-{\bf \Delta}_\perp^2)
\nonumber\\
\int \frac{dx^-}{4\pi} e^{ip^+x^- x}
\left\langle P\!\!+\!\!\Delta,\! 
\uparrow\!\left| \bar{q}\!\left(\!0\!\right)\gamma^+
q\!\left(\!{x^-}\!\right)
\right| P,\!\downarrow\right\rangle
&=& -\frac{\Delta_x\!\!-\!i\Delta_y}{2M}E(x,\!0,\!-{\bf \Delta}_\perp^2) 
\eea
Since $E$ is off diagonal in helicity, it will 
therefore only have a nonzero expectation value
in states that are {\sl not} eigenstates of helicity,
i.e. if we search for a probabilistic interpretation
for $E(x,0,t)$ we need to look for it in states that
are superpositions of helicity eigenstates.
For this purpose, we consider the state
\be
\left|X\right\rangle\equiv 
\left|p^+,{\bf R_\perp}=0,X\right\rangle
\equiv \left(
\left|p^+,{\bf R_\perp}=0,\uparrow\right\rangle
+
\left|p^+,{\bf R_\perp}=0,\downarrow\right\rangle
\right)/\sqrt{2}.
\ee
 In this state, we find
for the (unpolarized) impact parameter dependent
PDF
\bea
q_{X}(x,\!{\bf b_\perp}) \equiv
\left\langle X\left|O_q(x,{\bf b_\perp})\right|X
\right\rangle
= \!\!\int \!\!
\frac{d^2{\bf \Delta}_\perp}{(2\pi)^2}\! 
\left[ H_q(x,0,\!-{\bf \Delta}_\perp^2)
+ 
\frac{i\Delta_y}{2M} E_q(x,0,\!-{\bf \Delta}_\perp^2)
\right] \!e^{-i{\bf b}_\perp\cdot{\bf \Delta}_\perp}
%\nonumber
\eea
Upon introducing the Fourier transform of $E_q$
\be
{\cal E}_q(x,{\bf b_\perp})\equiv \int \frac{d^2
{\bf \Delta_\perp}}{(2\pi)^2}
 E_q(x,0,\!-{\bf \Delta}_\perp^2)
e^{-i{\bf b}_\perp\cdot{\bf \Delta}_\perp} 
\ee
we thus conclude that {\it $E_q$ describes how the
unpolarized PDF in the $\perp$ plane gets
distorted when the nucleon target is polarized
in a direction other than the $z$ direction}
\be
q_{X}(x,\!{\bf b_\perp}) =
q (x,\!{\bf b_\perp}) + \frac{1}{2M} 
\frac{\partial}{\partial b_y}
{\cal E}_q(x,{\bf b_\perp}).
\label{eq:distort}
\ee
The fact that the distorted distribution is still positive
implies further positivity constraints \cite{mb2,pub}.
Above distortion (\ref{eq:distort}) also shifts the 
$\perp$CM of the partons in the 
$y$-direction
\be
\langle x_q b^q_y\rangle \equiv
\int \!dx \!\int \!\! d^2{\bf b_\perp} x b^q_y
q_{X}(x,\!{\bf b_\perp})
= - \int\! dx\! \int\!\! d^2{\bf b_\perp} x 
\frac{{\cal E}_q(x,{\bf b_\perp})}{2M}
= - \int \! dx\, x \frac{E_q(x,0,0)}{2M},
\ee
i.e. the second moment of $E_q$ describes how
far the $\perp$CM is displaced transversely in a 
state with $\perp$ polarization (note that the
direction of the displacement is perpendicular to
both the $z$ axis and the direction of the
polarization. Of course, the net
(summed over all quark flavors plus glue)
displacement of the $\perp$CM
vanishes 
$\sum_{i\in q,g} \langle x_i b^i_y\rangle =0$
\cite{anom}.

The $\perp$ dipole moment due to
the displacement is given by
\be
d_q^y\equiv
\int\! dx\! \int\!\! d^2{\bf b_\perp} b_y
q_{X}(x,\!{\bf b_\perp})
= -\int \!\!dx\! \int \!\!d^2{\bf b_\perp} 
\frac{{\cal E}_q(x,{\bf b_\perp})}{2M}
= -
\int \!\!dx \frac{E_q(x,0,0)}{2M} = -\frac{\kappa_q(0)}{2M}
\ee
where $e_q\kappa_q$ is the contribution from flavor
$q$ to the anomalous Dirac moment $F_2(0)$. 
In order to get some feeling for the order of
magnitude, we consider a very simple model where 
only $q=u,d$ contribute to
$F_2(0)$, one finds for example $\kappa_d\approx -2$
and therefore a mean displacement of $d$ quarks of
by about $0.2fm$. For $u$ quarks the effect is about
half as large and in the opposite direction.

\begin{figure}
\unitlength1.cm
\begin{picture}(8,16)(2,5)
\includegraphics{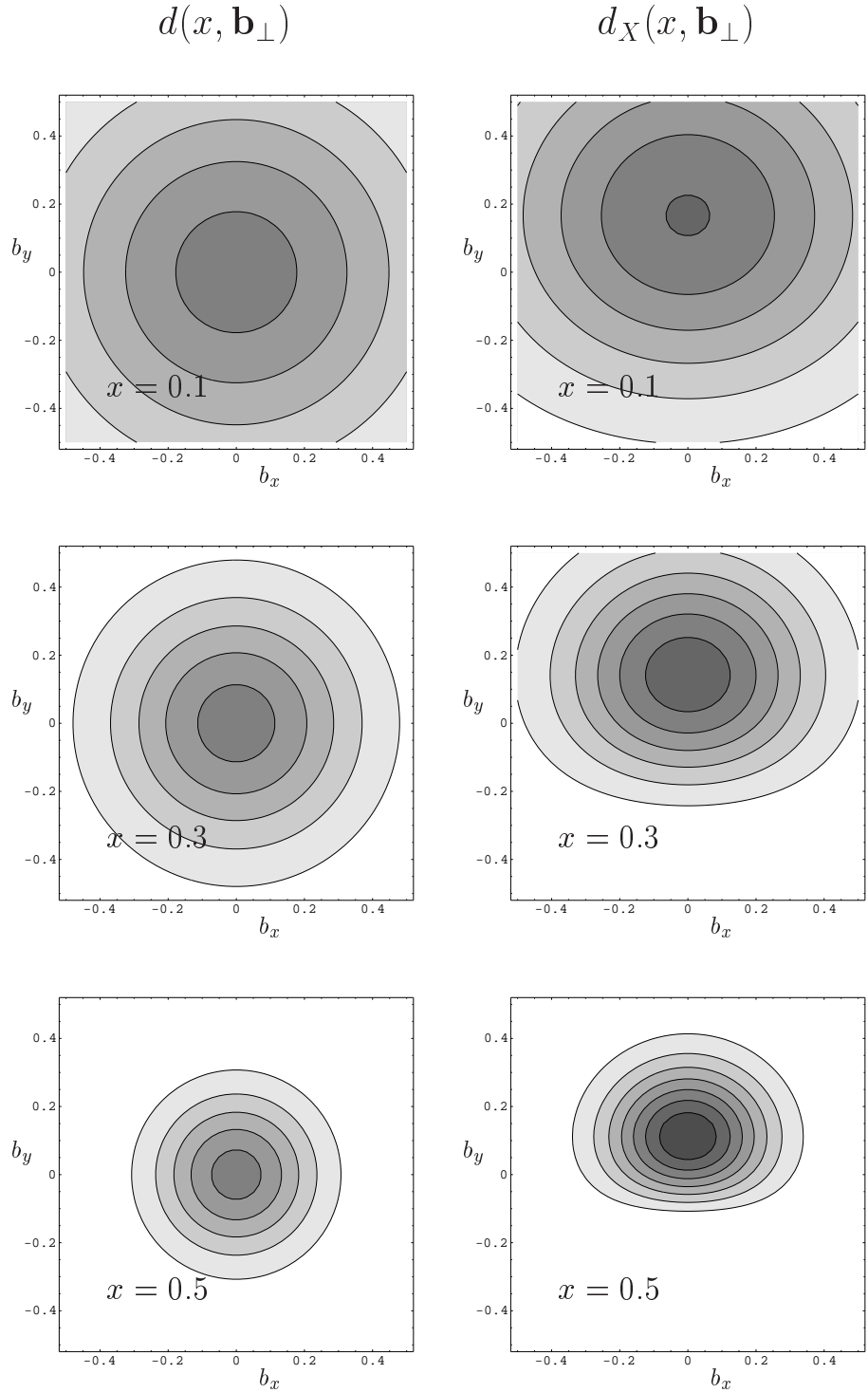}
\end{picture}
\caption{Impact parameter dependent PDF for $d$ quarks
$d(x,{\bf b_\perp})$  for $x\!=\!0.1$, $0.3$, $0.5$.
Left column: unpolarized;
right column: $d_X(x,{\bf b_\perp})$
in `$\perp$ polarized' proton .
The distributions are normalized to the central value
$d(x,{\bf 0_\perp})$.}
\label{fig:paneld}
\end{figure}

As a further illustration, we extend the model from
the previous section to $E_q(x,0,t)$, by making the
ansatz [the factor $\frac{1}{2}$ accounts
for $\int\! dx H_u(x,0,0)=2$]
\be
E_u(x,0,t) = \frac{1}{2}\kappa_u H_u(x,0,t)
\quad \quad \quad \quad 
E_d(x,0,t) = \kappa_d H_d(x,0,t).
\ee
Results 
for $d$-quarks are shown in Fig.
\ref{fig:paneld}.

\end{document}